\documentclass[secnumarabic,%
aps,%
pra,amsfonts,%
twocolumn,%
showkeys,floatfix%
]{revtex4-1}

\usepackage{bm,amsfonts}
\usepackage{sidecap}
\usepackage{placeins}
\usepackage{multirow}
\usepackage{graphicx}

\newcommand{\lafeasoA}{lafeaso1_arxiv}
\newcommand{\lafeasoB}{lafeaso2_arxiv}

\begin{document}
\title{Structural distortion as prerequisite for superconductivity in LiFeAs}  
\author{Ekkehard Kr\"uger}
\author{Horst P. Strunk}
\affiliation{Institut f\"ur Materialwissenschaft, Materialphysik,
  Universit\"at Stuttgart, D-70569 Stuttgart, Germany}
%
\date{\today}
\begin{abstract}
  The nonadiabatic Heisenberg model predicts a structural distortion in LiFeAs
  below a temperature higher than (or at least equal to) the superconducting
  transition temperature. Within this group-theoretical model, the reduction
  of the symmetry caused by the distortion is a prerequisite for the
  superconducting state in this compound and can be realized by a mere
  displacement of the iron atoms from their positions in the space group
  $P4/nmm$.
\end{abstract}
\maketitle

\section{Introduction}
LiFeAs becomes superconducting at the ``respectable'' transition temperature
$T_c = 18 K$ without any chemical doping~\cite{tapp}. In this way LiFeAs
differs remarkably from other high-$T_c$ FeAs compounds such as LaFeAsO which
becomes superconducting only on (slight) doping with fluorine (or other
specific dopants~\cite{izyumov}). As a pure substance, however, LaFeAsO
undergoes an antiferromagnetic spin-ordering transition at $\sim\!
137K$~\cite{clarina, nomura,kitao,nakai}.

The magnetic and superconducting properties of LaFeAsO have been analyzed
recently on the basis of the group-theoretical nonadiabatic Heisenberg model
(NHM)~\cite{\lafeasoA,\lafeasoB}. We reported evidence that both the
antiferromagnetic and the superconducting state are connected with a
magnetic~\cite{ea,\lafeasoA} and a superconducting~\cite{es1,\lafeasoB} band,
respectively, of special symmetry. As an important consequence, it is the {\em
  reduction of the symmetry} of LaFeAsO, caused in this case by the doping,
which activates superconductivity in this compound.

In the present paper we apply the NHM to LiFeAs and show, first, that the
magnetic order observed in LaFeAsO is not stable in LiFeAs and, second, that
in LiFeAs just as in LaFeAsO a superconducting state can be stable only when
the symmetry of the crystal is reduced. In the case of LiFeAs, this reduction
can be realized by a (small) displacement of the Fe atoms.

\section{Comparison of the band structures of
  L\lowercase{i}F\lowercase{e}A\lowercase{s} and
  L\lowercase{a}F\lowercase{e}A\lowercase{s}O}

LiFeAs and LaFeAsO possess common properties facilitating the comparison of
their band structures: both compounds possess the space group
$P4/nmm$~\cite{tapp,clarina,kamihara,prl_chen,nature_chen,wen,dong}, the Fe
atoms have the same positions in the unit cell, and the positions of the Li
and As atoms in LiFeAs are group-theoretically equivalent to the positions of
the La and As atoms in LaFeAsO. Solely the oxygen atoms are absent in
LiFeAs. As a consequence, magnetic and superconducting bands in both compounds
have the same symmetry as long as they are not related to the oxygen atoms. 

\begin{figure}[!]
  \includegraphics[width=.85\textwidth,angle=0]{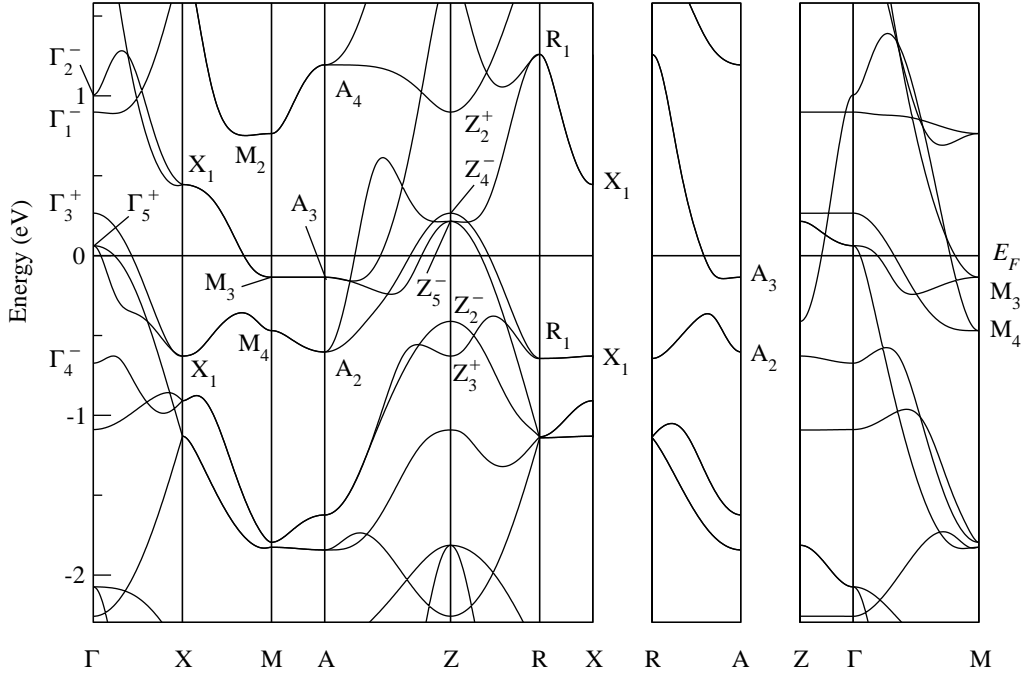}
  \caption{Band structure of tetragonal LiFeAs as calculated by the FHI-aims
    program \protect\cite{blum1,blum2}. The symmetry labels are determined by
    the authors and defined in Ref~\protect\cite{\lafeasoA}.   
  \label{fig:bandstr}
}
 \end{figure}

Comparing the band structure of LiFeAs as depicted in Fig.~\ref{fig:bandstr}
with the band structure of LaFeAsO (see Fig.~1 of Ref.~\cite{\lafeasoA}
or~\cite{\lafeasoB}), we find both an important difference and analogies: On
the one hand, LiFeAs does not possess a magnetic band related to the space
group $Imma$ as it exists in the band structure of LaFeAsO
(Sec.~\ref{sec:m_state}). On the other hand, both materials lack a
superconducting band in the space group $P4/nmm$ (Sec.~\ref{sec:s_state}).

\subsection{The magnetic order observed in LaFeAsO is not stable in LiFeAs} 
\label{sec:m_state}
In stoichiometric LaFeAsO the magnetic band related to the space group $Imma$
is evidently responsible for its observed~\cite{clarina, nomura,kitao,nakai}
antiferromagnetic state~\cite{\lafeasoA}. In the band structure of LiFeAs,
however, such a magnetic band {\em does not exist.} The reason is because the
Bloch function with $Z_1$ symmetry disappears in the band structure of
LiFeAs. This $Z_1$ state near the Fermi level, however, is an indispensable
component of the magnetic band in LaFeAsO~\cite{\lafeasoA}. Hence, the
symmetry of the Bloch functions of LiFeAs near the Fermi level is not
compatible with the magnetic structure experimentally determined in LaFeAsO
and, consequently, the magnetic order observed in LaFeAsO does not develop in
LiFeAs. We cannot exclude, however, that {\em other} magnetic structures with
other magnetic groups could be stable in distorted LiFeAs.

\subsection{Absence of a superconducting state in undistorted LiFeAs}
\label{sec:s_state}
Undistorted LaFeAsO lacks a superconducting band because the two branches of
any superconducting band in the space group $P4/nmm$ are degenerated at point
$M$ and are labeled by two representations $\Gamma^+$ and $\Gamma^-$ different
with respect to the inversion~\cite{\lafeasoB}. Hence, the run of the bands on
the lines $\Gamma X$ and $X M$ near the Fermi level of LaFeAsO and the
symmetry of the Bloch functions at point $\Gamma$ do not allow the
construction of a superconducting band in LaFeAsO~\cite{\lafeasoB}.

In LiFeAs, the run of the bands on the lines $\Gamma X$ and $X M$ near the
Fermi level is similar to the run of the related bands in LaFeAsO. In
addition, the Bloch functions of these bands possess the same symmetry as the
related Bloch functions in LaFeAsO.  As a consequence, also in the band
structure of LiFeAs we cannot detect a superconducting band and, hence, also
LiFeAs cannot develop a stable superconducting state in the space group
$P4/nmm$.

\section{Distorted L\lowercase{i}F\lowercase{e}A\lowercase{s} can exhibit
 superconductivity}
\label{sec:distorted}

There is now both a correspondence and an apparent contradiction between the
experimental observations and the predictions of the NHM: on the one hand,
LiFeAs does not develop the magnetic order observed in LaFeAsO as it is, in
fact, not allowed within the NHM. On the other hand, undoped LiFeAs shows
superconductivity, but theoretically the symmetry of the Bloch functions in
the band structure of LiFeAs does not permit a superconducting state in the
space group $P4/nmm$. However, this contradiction can be healed by a (small)
distortion of the crystal structure reducing the symmetry of $P4/nmm$.

In fact, several distortions of LiFeAs would allow a stable superconducting
state in the framework of the NHM. The chemical analogy of LiFeAs to LaFeAsO,
however, suggests that also in LiFeAs the symmetry of the system is reduced by
a (slight) displacement of the Fe atoms as it was experimentally observed in
LaFeAsO~\cite{clarina}. Indeed, the displacement of the Fe atoms as depicted
in Fig.~\ref{fig:structures} (a) as well as the displacement depicted in
Fig.~\ref{fig:structures} (b) changes the symmetry of the Bloch functions in
such a way that a stable superconducting state becomes possible.  The
displacement in Fig.~\ref{fig:structures} (a) realizes the space group $Pmm2$
(25) leaving the translation symmetry unchanged but reducing the point group
$D_{4h}$ of $P4/nmm$ to $C_{2v}$. The displacement depicted in
Fig.~\ref{fig:structures} (b), on the other hand, realizes the space group
$P4_2/nmc$ (137) leaving the point group unchanged but modifies the
translation symmetry of the crystal. The superconducting band related to
Fig.~\ref{fig:structures} (a) was already considered in
Ref.~\cite{\lafeasoB}. The superconducting band related to the displacement
depicted in Fig.~\ref{fig:structures} (b) is not published thus far. We shall
identify it once the displacement in Fig.~\ref{fig:structures} (b) is
experimentally confirmed.

\section{Summary }
\begin{figure}[t]
\begin{minipage}[c]{.18\textwidth}
\centering
\includegraphics[width=.7\textwidth,angle=0]{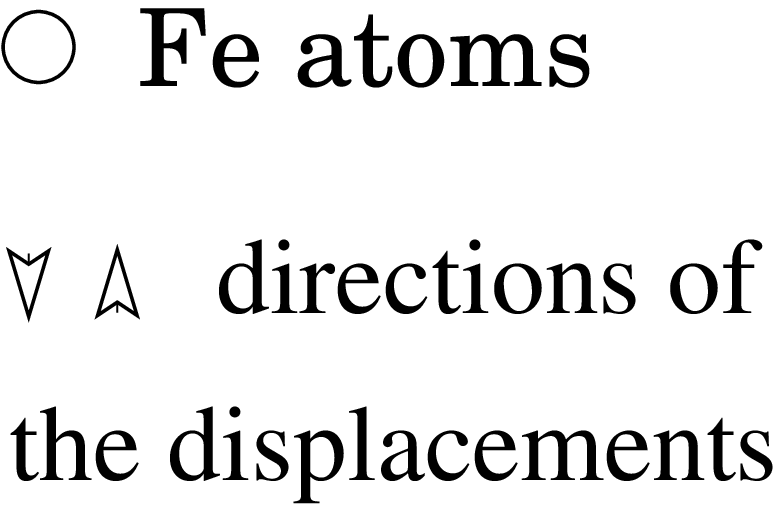}
\end{minipage}
\hspace{2cm}
\begin{minipage}[c]{.25\textwidth}
 \includegraphics[width=.93\textwidth,angle=0]{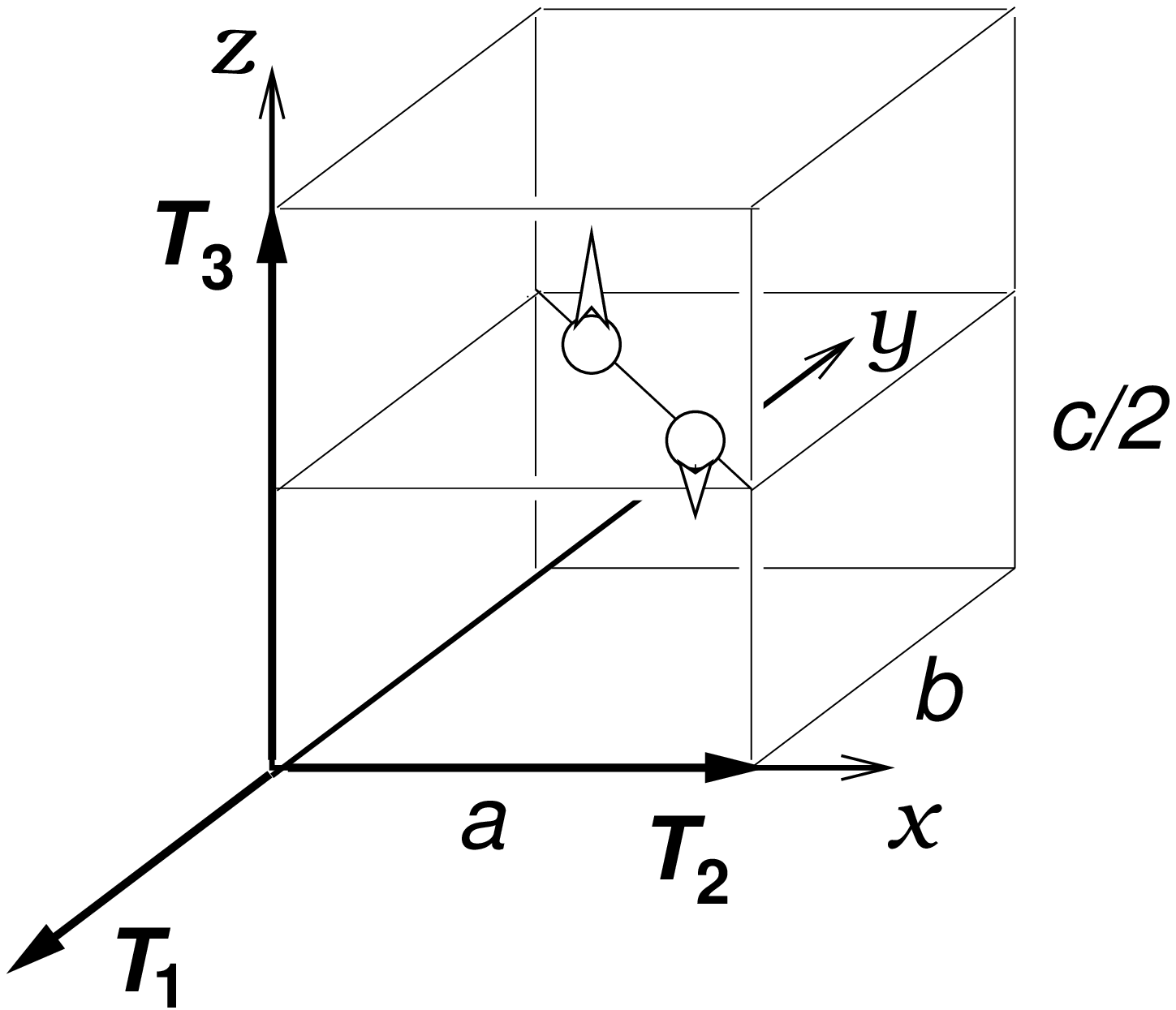}%
(a)
\end{minipage}
\hspace{2cm}
\begin{minipage}[c]{.21\textwidth}
\includegraphics[width=.93\textwidth,angle=0]{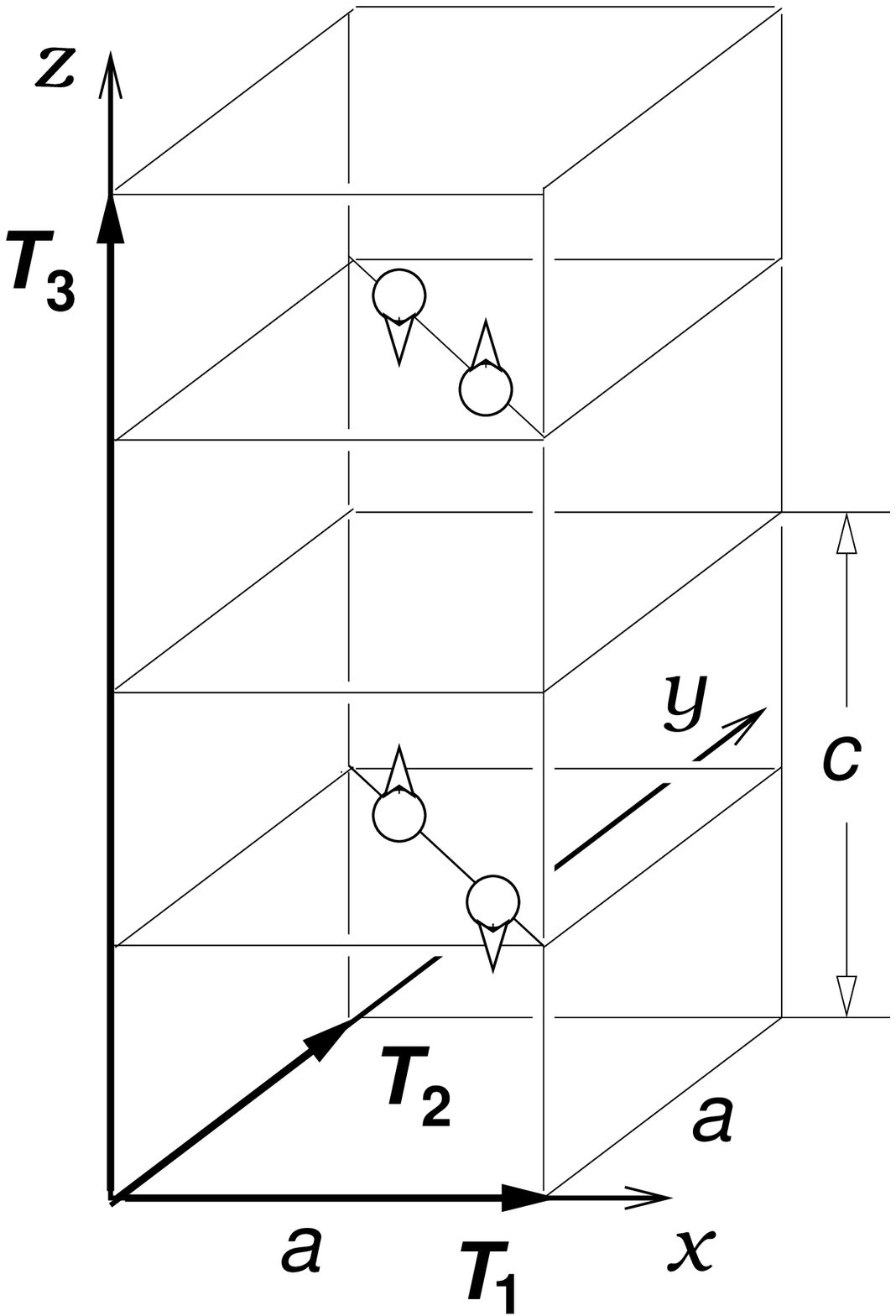}%
(b)
\end{minipage}
\caption{
Proposed unit cells and lattice vectors of two distorted LiFeAs crystals allowing  a
stable superconducting state. Both distortions are produced by a mere
displacement of the Fe atoms. For reasons of clarity, only the Fe atoms are
shown.  $a$, $b = a$, and $c$ stand for the lengths of the unit cell of
tetragonal undistorted LiFeAs (with the space group $P4/nmm$), and 
$\bm T_1$, $\bm T_2$, and $\bm T_3$ denote the basic translations 
of the corresponding Bravais lattice. 
(a) The displacement is periodic with the lattice
vector $T_3$ of undistorted LiFeAs and realize the orthorhombic space group
$Pmm2$ (25). The Fe atoms are displaced in $\pm z$ direction with different
amounts in the unit cell. 
(b) The unit cell of undistorted LiFeAs is doubled in $z$ direction. The Fe
atoms still are displaced in $\pm z$ direction, where a translation by the
lattice vector $T_3$ of undistorted LiFeAs effects an inversion of the
displacement. This displacement of the Fe atoms realizes the tetragonal space
group $P4_2/nmc$ (137) possessing the same point group $D_{4h}$ as the space
group $P4/nmm$ of undistorted LiFeAs. 
\label{fig:structures}
}
\end{figure}

We propose that LiFeAs undergoes a structural distortion below a temperature
$T_s$ higher than (or at least equal to) the superconducting transition
temperature $T_c$. The distorted system (below $T_s$) exhibits an interaction
between the electron spins and crystal-spin-1 bosons which produces below
$T_c$ stable Cooper pairs~\cite{\lafeasoB}. Several distortions in LiFeAs are
conceivable, the experimental observations~\cite{clarina} on LaFeAsO, however,
suggest that also in LiFeAs the distortion is realized by a displacement of
the Fe atoms as depicted in Fig.~\ref{fig:structures} (a) or (b).

\acknowledgements
  We are indebted to T\"unde Erdie from the H\"ochstleistungsrechenzentrum
  Stuttgart for kindly running the FHI-aims program and to Ernst Helmut Brandt
  for valuable discussion.

\onecolumngrid

\twocolumngrid

\FloatBarrier

\end{document}